\documentclass[11pt,twoside]{book}
\usepackage{konkolyproc2}
\usepackage{longtable}
\usepackage{amsmath,amssymb}
\usepackage{graphicx}
\usepackage{lscape}
\usepackage{index}
\usepackage{natbib}
\usepackage{bigdelim}
\usepackage{multirow}
\makeindex

\begin{document}

\pagestyle{myheadings}
\setcounter{equation}{0}\setcounter{figure}{0}\setcounter{footnote}{0}\setcounter{section}{0}\setcounter{table}{0}\setcounter{page}{1}
\markboth{Kanbur, Bhardwaj, Singh \& Ngeow}{RRL2015 Conf. Papers}
\title{Period-Color and Amplitude-Color Relations for RR Lyraes}
\author{Shashi Kanbur$^1$, Anupam Bhardwaj$^2$, H. P. Singh.$^2$, C. C. Ngeow$^3$, }
\affil{$^1$ SUNY Oswego, NY, USA 13126\\
$^2$University of Delhi, India\\
$^3$National Central University, Taiwan}

\begin{abstract}
We use published OGLE LMC/SMC data to present comprehensive Period-Color (PC) and Amplitude-Color (AC) relations for both fundamental and overtone stars. For fundamental mode stars, we confirm earlier work that the minimum light extinction corrected PC relation in V-I has a shallow slope but with considerable scatter (LMC: $[0.093 \pm 0.019]$ with a standard deviation about this line of 0.116, SMC: $[0.055\pm0.058]$ with a standard deviation about this line of 0.099). We note the high scatter about this line for both the LMC and SMC: either there is some source of uncertainty in extinction or some other physical parameter is responsible for this dispersion. We compare with previous results and discuss some possible causes for this scatter. In contrast, RRc overtone stars do not obey a flat PC relation at minimum light (LMC: $[0.604 \pm 0.041]$ with a standard deviation about this line of 0.109, SMC: $[0.472 \pm 0.265]$ with a standard deviation about this line of 0.091). The fact that fundamental mode RR Lyrae stars obey a flat relation at minimum light and overtone RR Lyrae stars do not is consistent with the interaction of the stellar photosphere and hydrogen ionization front. We compare these results with PC relations for fundamental and first overtone Cepheids. The fact that the PC relations change significantly as a function of phase indicates strongly that Cepheid and RR Lyrae relations can only be understood at mean light when their properties as a function of phase are determined. 

\end{abstract}

\section{Introduction}
RR Lyrae stars are very important objects because knowledge of their mean absolute magnitudes leads to a population II age and distance scale.
Their spatial distribution provides a way to map galaxy structure and hence constrain theories of galaxy formation. Here we study Period-Color (PC) and
Amplitude-Color (AC) relations as a function of phase. Applying the Stefan-Boltzmann law at the phases of maximum/minimum light we find
$$\log L_{max} - \log L_{min} \approx 4\log T_{max} - 4\log T_{min},\eqno(1)$$
where we make the assumption that radius fluctuations during a pulsation cycle are small, a reasonable assumption for
Cepheids and RR Lyraes (Cox 1980).

Equation (1) implies that if $T_{max}/T_{min}$ do not vary with period, then there will be a relationship between period and $T_{min}/T_{max}$ and
between amplitude and $T_{min}/T_{max}.$ We note that the observational counterpart of temperature is color.
Theoretical relations supporting a flat PC relation at {\it minimum} light for FU RR Lyraes are provided in Kanbur (1995) and Kanbur and Phillips (1996).

The hydrogen ionization front (HIF) and stellar photosphere (SP) are not co-moving as the star pulsates. In certain situations they can be engaged
with the SP lying at the base of the HIF. Then the color of the star is the temperature at which hydrogen ionizes.
The SP cannot go further inside the mass distribution of the star because of the large opacity associated
with hydrogen ionization. When the SP and HIF are engaged in certain envelope temperature/density ranges,
the temperature at which hydrogen ionizes and hence the color
of the star is somewhat independent of global stellar parameters such as period. This engagement varies with stellar period/pulsation phase and metallicity. 
For example, FO RR Lyraes are generally hotter than FU stars. This increased temperature means that the temperature at which hydrogen ionizes depends more 
strongly on period. Thus even though the SP and HIF are engaged, the color of the star depends more strongly on period at minimum light than is the case for FU
stars.
These same considerations imply that FO RR Lyraes should not follow a flat PC relation at minimum light because such stars are hotter than FU stars. 

\section{Results}
Figures 1 and 2 are adapted from B14 and present results for fundamental mode (FU, figure 1)/ first overtone (FO, figure 2)
LMC (top panels)/SMC (bottom panels) RR Lyraes observed by OGLE III (Soszynski et al., 2008)
published in Bhardwaj et al (2014, B14). The left/right panels are PC/AC relations. Minimum/maximum light are represented by blue/red points respectively.
The data were corrected for reddening and extinction using Haschke et al (2011) maps.
Tables 1 and 2 present these results quantitatively.

For FU stars, we clearly see a flat or flatter PC relation at minimum light with a significant relation at maximum. We also see that
RRc stars do not follow a flat relation at minimum light. These observations are consistent with the ideas presented in Kanbur and Phillips (1996).  
The AC relations are also broadly consistent with these ideas.

\begin{table}[!hb] 
\caption{Slope and Intercept for PC/AC relations at maximum/minimum light for FU stars.} 
\smallskip
\begin{center}
\begin{tabular}{lrrrr}
\tableline
\noalign {\smallskip} 
&Phase&Slope(RRab)&Intercept(RRab)&$\sigma (RRab)$\\ 
\noalign{\smallskip}
\tableline
\noalign{\smallskip}
\multicolumn{5}{c}{LMC}\\
PC&max&$1.505\pm0.018$&$0.654\pm0.004$&0.116\\
&min&$0.093\pm0.019$&$0.716\pm0.005$&0.116\\
AC&max&$-0.361\pm0.003$&$0.392\pm0.002$&0.091\\
&min&$0.049\pm0.003$&$0.631\pm0.003$&0.114\\
\multicolumn{5}{c}{SMC}\\
PC&max&$1.768\pm0.053$&$0.705\pm0.012$&0.097\\
&min&$0.055\pm0.058$&$0.725\pm0.013$&0.099\\
AC&max&$-0.370\pm0.007$&$0.594\pm0.006$&0.074\\
&min&$0.067\pm0.010$&$0.660\pm0.008$&0.098\\
\noalign{\smallskip}
\tableline
\noalign{\smallskip}
\end{tabular}  \end{center}
 
\end{table}

\begin{table}[!hb]
\caption{Slope and Intercept for PC/AC relations at maximum/minimum light for FO stars.}
\smallskip
\begin{center}
\begin{tabular}{ccrcc}
\tableline
\noalign {\smallskip}
&Phase&Slope(RRc)&Intercept(RRc)&$\sigma (RRc)$\\
\noalign{\smallskip}
\tableline
\noalign{\smallskip}
\multicolumn{5}{c}{LMC}\\
PC&max&$0.770\pm0.032$&$0.638\pm0.015$&0.084\\
&min&$0.604\pm0.041$&$0.836\pm0.020$&0.109\\
AC&max&$-0.089\pm0.014$&$0.312\pm0.007$&0.091\\
&min&$0.411\pm0.017$&$0.357\pm0.008$&0.111\\
\multicolumn{5}{c}{SMC}\\
PC&max&$0.536\pm0.228$&$0.529\pm0.228$&0.079\\
&min&$0.472\pm0.265$&$0.823\pm0.116$&0.091\\
AC&max&$-0.312\pm0.078$&$0.450\pm0.030$&0.074\\
&min&$0.172\pm0.098$&$0.530\pm0.050$&0.094\\
\noalign{\smallskip}
\tableline
\noalign{\smallskip}
\end{tabular}  \end{center}
\end{table}

\begin{figure}[!ht]
\includegraphics[width=1.0\textwidth]{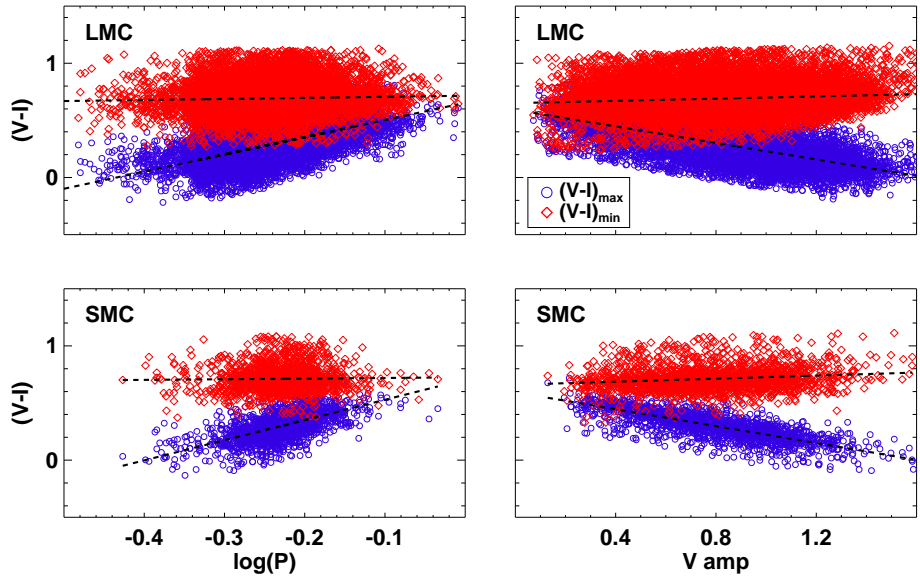}
\caption{RR Lyrae Fundamental mode Period-Color/Amplitude Color relations at maximum (blue)/minimum (red) light for
the SMC and LMC} 
\label{authorsurname-fig1} 
\end{figure}

\begin{figure}[!ht]
\includegraphics[width=1.0\textwidth]{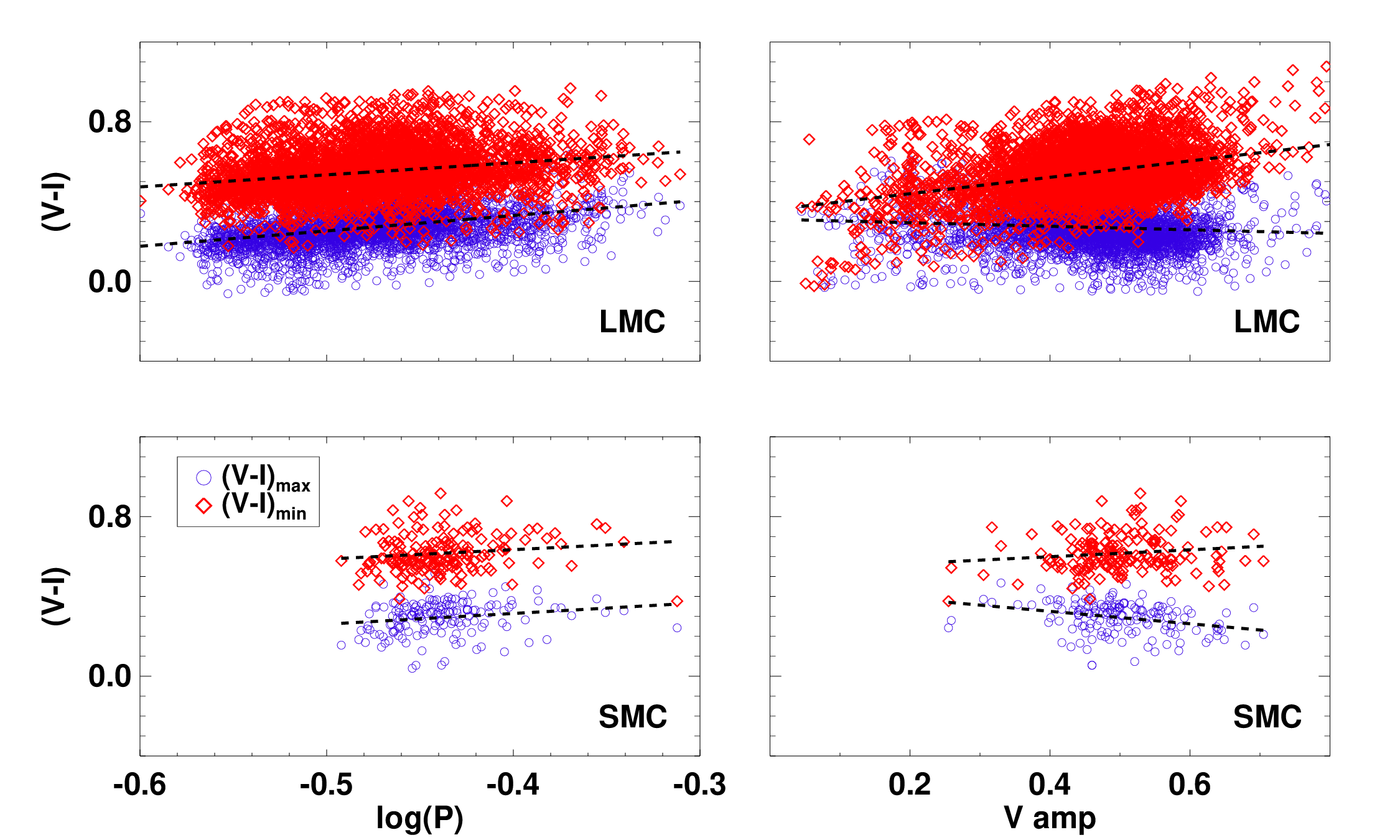}
\caption{RR Lyrae First Overtone Period-Color/Amplitude Color relations at maximum (blue)/minimum (red) light got the
SMC and LMC}
\label{authorsurname-fig1}
\end{figure}

\section{ Conclusion}

Our result are to be compared with those for Cepheids (B14 and references therein) that display a flat PC relation at {\it maximum} light.
Because Cepheids are cooler, the HIF lies further inside the mass distribution and hence the SP and HIF are only engaged at maximum light.

The broader implication is that Cepheid and RR Lyrae pulsation properties such as Period-Luminosity/Period-Wesenheit/Period-Color relations
are always quoted at mean light. However, figures 1 and 2 clearly demonstrate that such relations change as a function of pulsation phase: mean
light relations are obtained by averaging over the corresponding relations at different phases. We suggest that a
greater understanding of Cepheid and RR Lyrae pulsation will only occur if these relations are studied as a function of pulsation phase.

This work was supported by a grant from the Indo-US Science and Technology Forum and by a travel grant from SUNY Oswego.


\end{document}